\documentclass[%
 aip,
 amsmath,amssymb,
 reprint,%
]{revtex4-1}

\usepackage{graphicx}% Include figure files
\usepackage{dcolumn}% Align table columns on decimal point
\usepackage{bm}% bold math
\usepackage[utf8]{inputenc}
\usepackage[T1]{fontenc}
\usepackage{mathptmx}
\usepackage{etoolbox}

\makeatletter
\def\@email#1#2{%
 \endgroup
 \patchcmd{\titleblock@produce}
  {\frontmatter@RRAPformat}
  {\frontmatter@RRAPformat{\produce@RRAP{*#1\href{mailto:#2}{#2}}}\frontmatter@RRAPformat}
  {}{}
}%
\makeatother
\begin{document}

\preprint{AIP/123-QED}

\title[Extreme forcing and wave dynamics in coupled excitable FitzHugh–Nagumo systems]{Extreme forcing and wave dynamics in weakly nonlocally coupled excitable FitzHugh–Nagumo systems}
% Force line breaks with \\
\author{N.I. Semenova}%
 \email{semenovani@sgu.ru}
\author{V.V. Semenov}%
\author{A.V. Bukh}%
 \affiliation{Saratov State University, Astrakhanskaya str. 83, Saratov 410012, Russia}%

\date{\today}% It is always \today, today,

\begin{abstract}
The influence of extreme external forcing on traveling-wave dynamics in an ensemble of weakly nonlocally coupled excitable FitzHugh--Nagumo systems is studied. Three types of external exposure are considered: periodic Gaussian pulses, periodic pulses modulated by Gaussian white noise, and Lévy noise with tunable distribution parameters. Periodic forcing produces synchronization tongues with highly regular collective dynamics and may induce multiple traveling waves or coexistence of partial synchronization with wave propagation. In contrast, Lévy noise suppresses regular behavior and generates a regime of counter-propagating waves, which with increasing intensity transitions to random walking dynamics. The study provides a comprehensive classification of the observed dynamical regimes and presents their organization in parameter space for different types of extreme external forcing.
\end{abstract}

\maketitle

\begin{quotation}
Traveling waves arise in a wide spectrum of physical, chemical, biological, ecological and other spatially-extended systems, as well as in networked oscillators. Mathematical models that demonstrate such wave phenomena are extensively employed to describe a broad class of regular and chaotic spatiotemporal dynamics, including processes such as signal and energy transmission, climate and population changes, neural activity, propagation of infectious diseases and ocean waves, etc. Generation, suppression, and modulation of traveling waves can be achieved through varying the system parameters being responsible for local dynamics, by adjusting the properties of spatial interactons or by applying deterministic or stochastic external forcing. The last approach is discussed in the current paper on an example of an ensemble of coupled excitable units subject to various deterministic and stochastic signals. In particular, periodic Gaussian pulses, pulse exposure modulated by Gaussian white noise, and Lévy noise with adjustable parameters controlling the symmetry and mean of the distribution.
\end{quotation}

\section{Introduction}\label{sec:intro}
The term 'traveling wave' is incredibly general and encompasses diverse deterministic and stochastic processes occurring in media \cite{landa1996,ghazaryan2022,smoller1994,garcia1999}, networks \cite{korneev2021,korneev2022,semenov2023,zakharova2023,rybalova2023,semenov2025-2}, single and coupled oscillators with long delay considered as spatially-extended systems \cite{giacomelli2012,klinshov2017,semenov2018,zakharova2024,semenov2025,semenov2026}. Traveling waves associated with spatially-periodic patterns, propagating fronts, backs and pulses are found to occur in plasma physics \cite{loarte1999,schwabe2007}, optics and electronics \cite{ebeling1993,akhmediev1997,semenov2023-2,liu2016,semenov2025,semenov2026}, hydrodynamics \cite{henry2019,debnath1994,johnson1997,dias1999}, chemistry \cite{kuramoto1984,kapral1995,epstein1998}, neurophysiology \cite{hugles1995,muller2018,mohan2024}, as well as in interdisciplinary fields at the intersection of sciences, such as ecology, population biology and epidemiology \cite{sherratt2008,malchow2008,brauer2012}. Traveling waves continue to attract attention within the fields of nonlinear dynamics and theory of complex systems due to the wide scope of dynamical systems, processes  and related applications. In particular, special attention is drawn to the problem of identifying universal properties of traveling waves and developing approaches for controlling their behavior and stability.

The present research addresses the traveling wave dynamics in networked excitable oscillators. There are several strategies for control (i.e. acceleration, deceleration, suppression, or generation) of traveling waves in such systems. The first group of methods involves the properties of spatial interaction and coupling. In particular, memristive coupling (a particular case of adaptive interaction) is shown to be an appropriate tool for controlling traveling waves in single-layer \cite{korneev2021,korneev2024-2} and multilayer \cite{korneev2022} networks of excitable nodes. A suppressive character of higher-order interactions on spiral waves in ensembles of excitable oscillators is reported in Ref. \cite{hu2024}. In excitable media, one can use the propetries of nonlocal coupling of various spatial kernel for the traveling wave control \cite{bachmair2014,dahlem2008,schneider2009}. Alternatively,  time-delayed feedback \cite{dahlem2008,schneider2009} can be applied for this purpose. Initially reported on examples of excitable media, the approaches based on nonlocal interaction and time-delayed feedback can be similarly realized in ensembles of coupled oscillators. The third set of methods refers to control by means of external forcing and adjusting its parameters. Such impacts can be pure deterministic \cite{steinbock1993,shepelev2016}, as well as contain both deterministic and stochastic components \cite{yuan2011}. In the presence of pure stochastic  influence, one can vary its characteristics to control the processes of wave generation and suppresion  as demonstrated in Ref. \cite{korneev2024-2} on an example of Lévy noise. 

Developing the concept of traveling wave control in excitable systems via external forcing, the current paper is focused on control by means of impulsive signals, both deterministic, with tunable amplitude and frequency, and stochastic, whose amplitude is noise-modulated. As the stochastic component, both Gaussian and Lévy noises are taken into account. It is shown that impulsive perturbations enable not only the induction and suppression of traveling waves, as in the case of additive Gaussian and Lévy noise \cite{korneev2024-2}, but also the realization of more complex hybrid regimes that simultaneously reflect the traveling wave dynamics and other phenomena, such as the formation of complex spatiotemporal structures and synchronization. Thus, the study simultaneously addresses the problems of traveling waves, Lévy-noise-induced effects, synchronization and pattern formation.

\section{Model and methods}\label{sec:model}

In this paper, we study the spatiotemporal effects occurring in an ensemble of the identical FitzHugh-Nagumo oscillators in the excitable regime depending on the external influence $I(t)$. The partial elements are connected by weak nonlocal coupling, thus forming a ring of oscillators. The system of equations describing the system under study are given below:
\begin{equation}\label{eq:FHN}
\begin{array}{c}
\varepsilon \frac{du_i}{dt}=u_i - \frac{u_i^3}{3}-v_i+
\frac{\sigma}{2R}\sum\limits^{i+R}_{j=i-R} (u_j-u_i) + I(t) , \\
\frac{dv_i}{dt}=u_i+a_i,
\end{array}
\end{equation}
where $u_i$ and $v_i$ are the activator and inhibitor variables, respectively, the oscillator's number is set by $i=1,...,N$ with $N=500$ being the total number of elements in the network with periodic boundary conditions. A small parameter $\varepsilon>0$ sets the time scale separation of fast activator and slow inhibitor variables, while $a$ defines the excitability threshold. For each partial FHN system, it determines whether the system is in the excitable ($|a|>1$), or oscillatory ($|a|<1$) regime.

The parameter $R$ indicates the number of nearest neighbours in each ring direction coupled with $i$th element. For convenience, we additionally introduce the coupling range set by $r=R/N$ . The strength of the coupling is characterized by $\sigma$. For our simulations, the initial conditions randomly distributed on circle $u^2+v^2\le 2^2$. At the same time, the maps of regimes require specially prepared initial conditions with a stabilized regime, in order to obtain the possible boundaries. 

In the system under study, the external influence is designated as $I(t)$. Further, we will use it for introducing the external influence: periodic pulse exposure, noisy periodic pulses and Lévy noise.

\begin{figure}[h]
\includegraphics[width=0.7\linewidth]{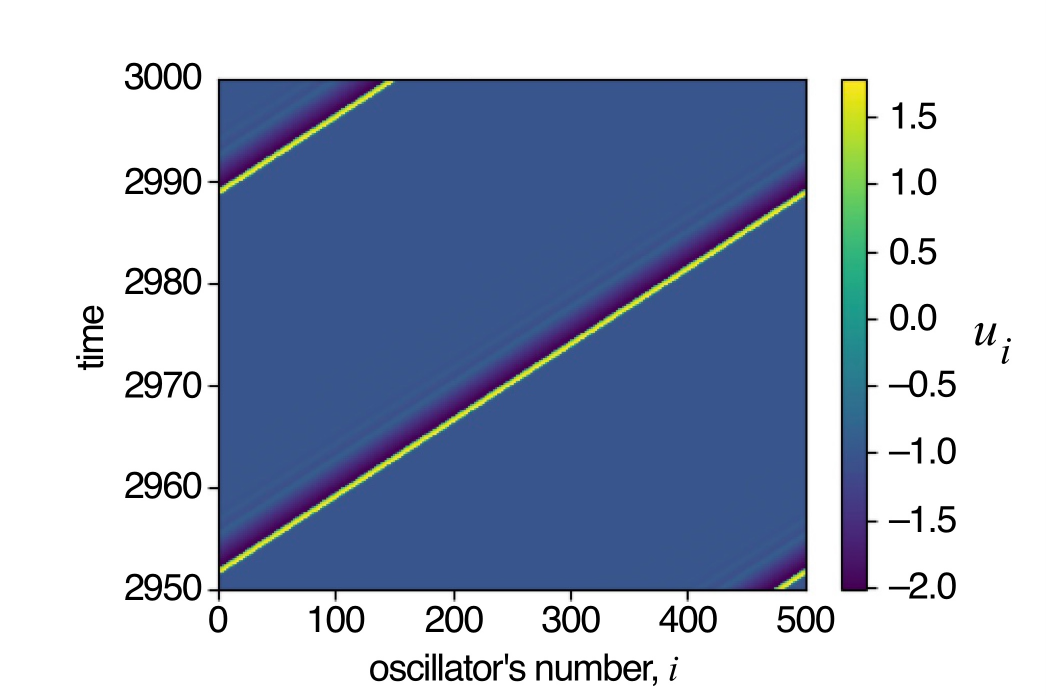}
\caption{\label{fig:trav_wave} A single traveling wave stabilized in the ensemble (\ref{eq:FHN}) of weakly nonlocally coupled FitzHugh--Nagumo systems in excitable regime in the absence of external forcing, i.e., for $I(t)=0$. Parameters: $\varepsilon=0.05$, $a=1.01$, $\sigma=0.2$, $r=0.006$.}
\end{figure}

The model was integrated using the Heun's method (the modified Euler method) with a dimensionless time step of $h = 10^{-3}$. In numerical experiments involving Lévy noise, the integration step was adaptively reduced when necessary. An additional clarification regarding this procedure will be included in Section \ref{sec:levy}. The remaining parameters of the individual FitzHugh--Nagumo subsystems were fixed as follows: $\varepsilon = 0.05$ and $a = 1.01$, corresponding to the excitable regime. Regarding the coupling between the partial elements, the coupling strength was set to $\sigma = 0.2$, while the coupling radius was chosen as $r = 0.006$, meaning that each element in the ensemble was coupled to its three nearest neighbors on either side demonstrating a weak nonlocal coupling. This parameter choice ensures that, for random initial conditions, the autonomous ensemble converges in most cases to a stable equilibrium state without oscillatory activity. However, in rare cases, the system may develop a single traveling wave, as illustrated in Fig.~\ref{fig:trav_wave}.

The parameters specified above remain fixed throughout the paper. The main control parameters considered in this study are the characteristics of the external forcing, including the amplitude and period of the periodic external force, as well as the intensities and other parameters of the stochastic perturbations.

\section{Periodic pulse exposure}\label{sec:pulses}
In order to introduce a pulse exposure and maintain the integrability of the system, here we consider Gaussian pulses shaped as a Gaussian function \cite{Chomycz2009,Bauer1984}: 
\begin{equation}\label{eq:one_pulse}
P(t)=A\cdot\exp\Big(-\frac{t^2}{2\Delta t^2}\Big),
\end{equation}
where $A$ is the amplitude of pulse, parameter $\Delta t$ controls the width of Gaussian pulse. Its smaller values make the pulse narrower. At the same time, it is important to ensure that when integrating a system with pulse exposure, the integration step is sufficient to adequately simulate pulses, so the pulses should not be made too narrow.

In this paper we are mainly interested in pulse-periodic exposure. Equation~\ref{eq:one_pulse} describes only one impulse. In order to make the influence periodic, it is necessary to replace $t$ with a periodic function of $t$:
\begin{equation}\label{eq:periodic_pulses}
I(t)=A\cdot\exp\Big(-\frac{1}{2 s} \sin^2\frac{w t}{2}\Big).
\end{equation}
This is how the external pulse-periodic exposure $I(t)$ is set in (\ref{eq:FHN}). It describes periodic pulses with amplitude $A$, frequency $w$ and pulse width controlled by the parameter $s$. Further, we will consider the impact of external influence depending on its amplitude $A$ and period $T=2\pi/w$.

In the ensemble (\ref{eq:FHN}), the external pulse exposure applied to each neuron $i$ evolves identically in time according to Eq.~(\ref{eq:periodic_pulses}).

\begin{figure}[t]
\includegraphics[width=\linewidth]{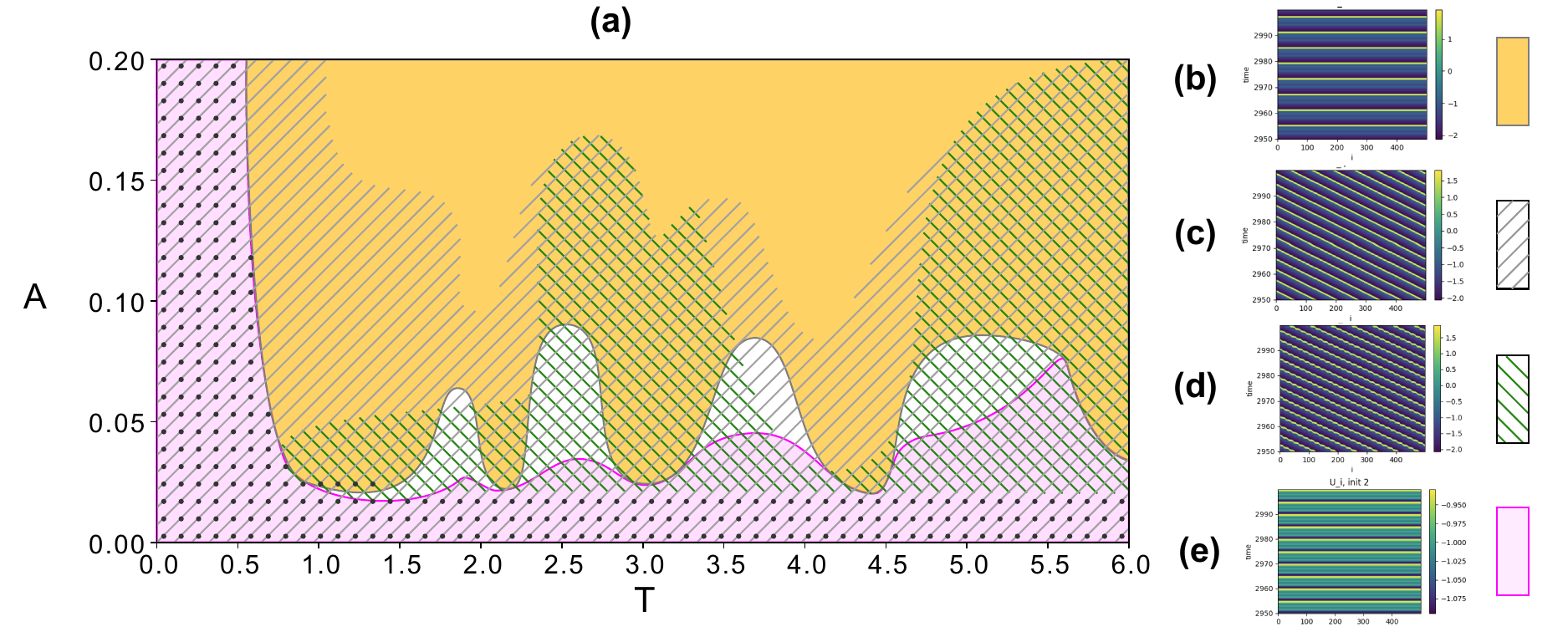}
\caption{\label{fig:pulse1} Regime diagram on the parameter plane $(A, T)$ of periodic pulse exposure (a). The dotted region correspond to a single traveling wave while the rest regimes are indicated using corresponding space-time plots: complete spatial synchronization (b), multiple traveling waves (c), multiple oscillating traveling waves (d), weak oscillations around the equilibrium state (e).}
\end{figure}

Figure \ref{fig:pulse1} shows a phase diagram of the main dynamical regimes observed upon variation of the parameters of the pulse exposure (\ref{eq:periodic_pulses}), namely the amplitude $A$ and the period $T = 2\pi / w$. For small amplitudes and periods of the external forcing, only the same regimes as in the absence of external forcing are observed. Since the individual elements operate in the excitable regime, their dominant dynamical state is a stable fixed point of focus type without oscillatory activity. Weak periodic forcing induces small-amplitude oscillations around the equilibrium state, while no spiking activity is generated. This regime is shown by the pink region in Fig.~\ref{fig:pulse1}(a), and an example of the corresponding spatio-temporal diagram is presented in panel (f) of the same figure.

For certain initial conditions and the same parameter values, a traveling wave may form in the system, as previously shown in Fig.~\ref{fig:trav_wave}. This regime can also be observed under weak external forcing with small amplitude or short period. The corresponding region in the parameter space is indicated by the dotted area in Fig.~\ref{fig:pulse1}(a).

In the presence of external periodic forcing, the emergence of a regime of complete spatial synchronization is a natural outcome. An example of this regime is shown in Fig.~\ref{fig:pulse1}(b). It is characterized by the presence of synchronous spiking activity throughout the entire observation time. This regime is predominantly observed at large amplitudes of the external forcing, and the corresponding region in the parameter space is indicated by the orange area in Fig.~\ref{fig:pulse1}(a). A characteristic feature of these regions is the presence of so-called synchronization tongues, where synchronization is achieved at lower forcing amplitudes and occurs at rational frequency ratios. For example, in Fig.~\ref{fig:pulse1}(a), synchronization tongues are observed at forcing periods approximately equal to $1.1$, $2.2$, $3.3$, and $4.4$. A similar effect has been reported in the study of the influence of pulsed forcing on chimera states in Ref.~\cite{Rybalova2024}.

\begin{figure}[t]
\includegraphics[width=\linewidth]{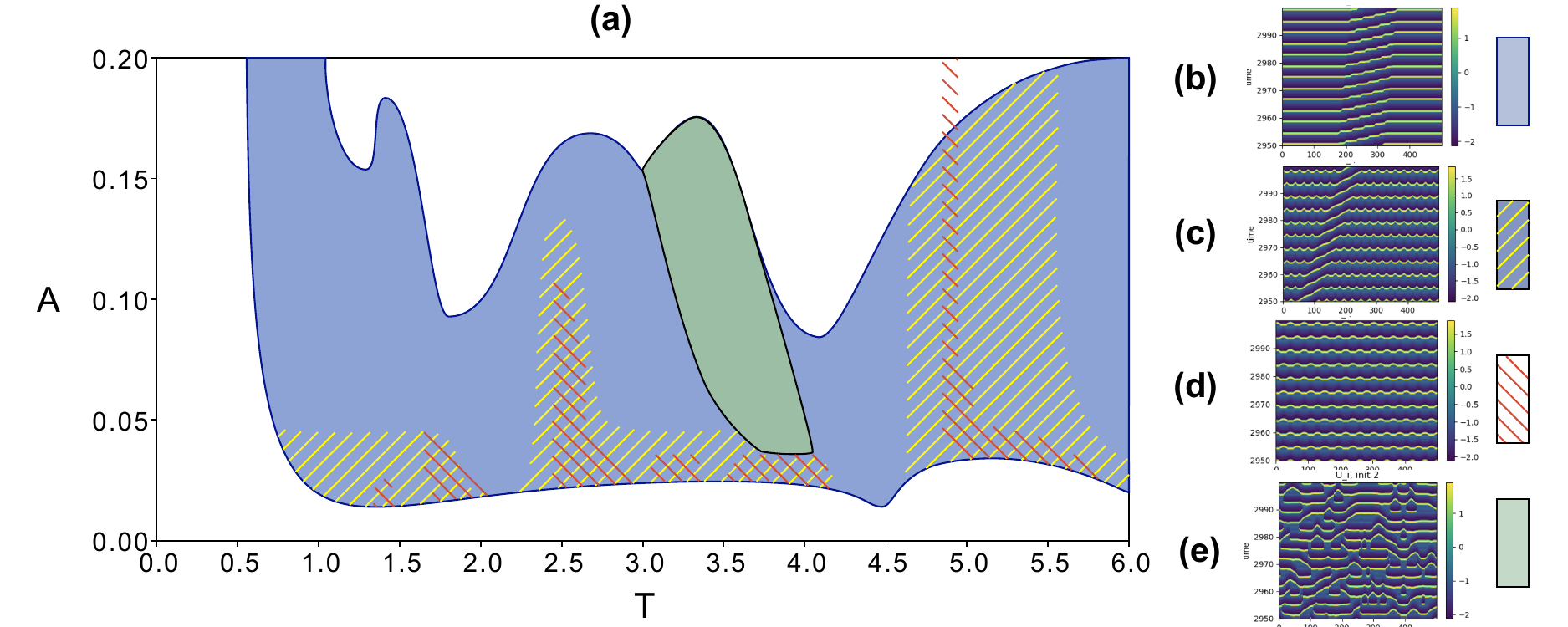}
\caption{\label{fig:pulse2} Regime diagram on the parameter plane $(A, T)$ of periodic pulse exposure (a). All regimes are indicated using corresponding space-time plots: coexistence of partial spatial synchronization and traveling waves (b), coexistence of standing and traveling waves (c), standing waves (d), randomly walking waves (e).}
\end{figure}

As the period of the external forcing is varied, new spatiotemporal regimes begin to emerge. Figure \ref{fig:pulse1}(a) shows the most representative of these derived regimes. For instance, when the forcing period is increased, a single traveling wave is replaced by multiple coexisting waves. The corresponding spatiotemporal dynamics is shown in Fig.~\ref{fig:pulse1}(c), while the region of existence of this regime is indicated by the gray hatched area in Fig.~\ref{fig:pulse1}(a). A similar regime, but characterized by weak oscillations, is presented in Fig.~\ref{fig:pulse1}(d) and marked by the green hatched region. These two regimes coexist and can be realized for different initial conditions. In fact, this property holds more generally across the system: the ensemble exhibits strong multistability, so that different initial conditions can lead to different dynamical regimes. In the intermediate region between complete spatial synchronization and the initial oscillatory or single-traveling-wave states, a wide variety of complex regimes is observed. While Fig.~\ref{fig:pulse1}(a) highlights regimes more closely related to traveling-wave dynamics, Fig.~\ref{fig:pulse2}(a) focuses on regimes that are more closely associated with spatial synchronization.

One of these intermediate regimes, characterized by the coexistence of partial spatial synchronization and traveling waves, is shown in Fig.~\ref{fig:pulse2}(b). This regime can be observed over a relatively broad parameter range and is indicated by the blue region in Fig.~\ref{fig:pulse2}(a).

In a certain parameter region, partial spatial synchronization transforms into standing waves (Fig.~\ref{fig:pulse2}(c), yellow hatched area in Fig.~\ref{fig:pulse2}(d)). In this case, the two regimes involving partial traveling-wave structures exhibit almost no parameter overlap and therefore do not coexist. A similar transformation is also observed for the regime of full synchronization, which evolves into standing waves (Fig.~\ref{fig:pulse2}(d)). This regime can be easily obtained from random initial conditions at small amplitudes of the external forcing (red hatched region in Fig.~\ref{fig:pulse2}(a)). Interestingly, the same regime can also be realized at $T = 4.9$ over a broad range of $A$ values; however, in this case it is typically observed only for specially chosen initial conditions.

In addition to the large variety of regimes resembling spatial synchronization, traveling waves, and standing waves, the system also exhibits a regime characterized by spatial incoherence, shown in green in Fig.~\ref{fig:pulse2}(a). An example of the corresponding spatiotemporal diagram is presented in Fig.~\ref{fig:pulse2}(e).

All of the regimes described above were initially obtained using random initial conditions, where the initial values of the variables were uniformly distributed on the circle $u^2 + v^2 = 2^2$. However, in order to construct the regime diagram and accurately determine the boundaries of existence of the different regimes, specially prepared initial conditions corresponding to a given stationary regime were employed.

In the regime diagrams shown in Figs.~\ref{fig:pulse1} and \ref{fig:pulse2}, the dynamical regimes are presented for external forcing amplitudes up to $0.2$. At higher amplitudes, all emerging nontrivial regimes are destroyed, and only complete spatial synchronization persists. A similar behaviour was previously observed in the study of the effect of the same type of forcing on chimera states in Ref.~\cite{Rybalova2024}.

\section{Extreme exposure with white Gaussian noise}\label{sec:wgn}
To investigate which regimes are specifically induced by the periodic component while preserving the extreme nature of the forcing, we consider an input that combines features similar to the periodic pulse forcing studied above. At the same time, a stochastic contribution in the form of a Gaussian process is introduced:
\begin{equation}\label{eq:noisyPulses}
I(t)=\sqrt{2D} \xi(t) \cdot \exp\Big(-\frac{1}{2 s} \sin^2\frac{w t}{2}\Big).
\end{equation}
In this equation, the second term corresponds to the periodic pulse forcing introduced earlier; however, the forcing amplitude is not explicitly specified. Instead, a stochastic multiplier $\sqrt{2D},\xi(t)$ is introduced, where $\xi(t)$ denotes a Gaussian white noise source and $D$ is the noise intensity. Due to the multiplicative coupling with the stochastic term $\sqrt{2D},\xi(t)$, the amplitude of the pulses fluctuates randomly in time.

In the previous section, the periodic pulsed exposure $I(t)$ was considered, which varies in time but is identical for all oscillators in the ensemble. In contrast, for the case of extreme forcing with additive Gaussian white noise in (\ref{eq:noisyPulses}), the periodic component remains time-dependent, whereas the noise source generates independent realizations for different oscillators, i.e., $\xi_i(t)$.

\begin{figure}[t]
\includegraphics[width=\linewidth]{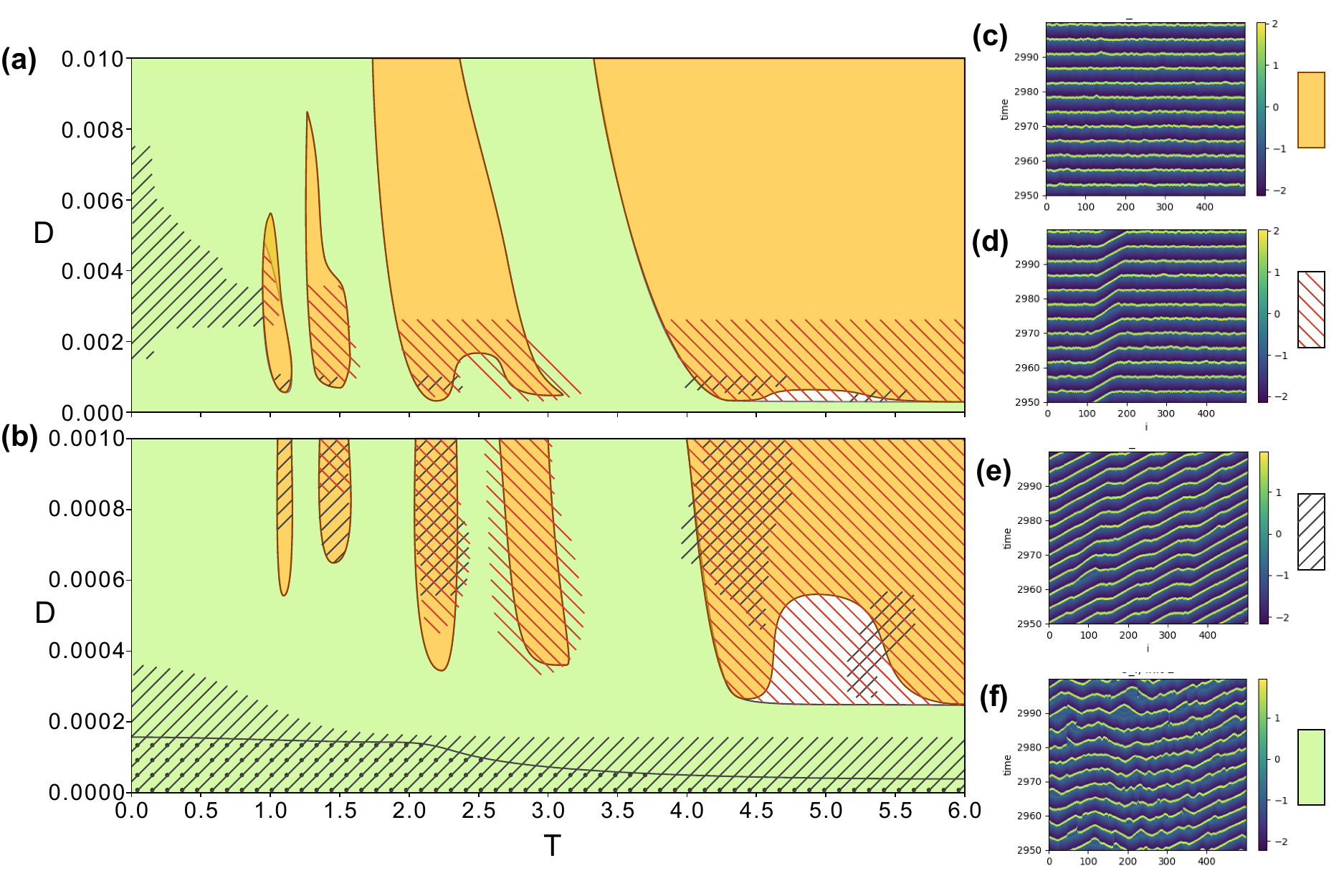}
\caption{\label{fig:wgn_uncorr} Regime diagram on the parameter plane $(D, T)$ of noisy periodic pulse exposure (a,b). The dotted region correspond to a single traveling wave coexisting with weak oscillations around the equilibrium state. The rest regimes are indicated using corresponding space-time plots: almost coherent spiking behaviour (c), partial coherent behaviour with partial traveling waves (d), multiple oscillating traveling waves (e), randomly walking waves (e).}
\end{figure}

Analogously to the case of periodic pulsed forcing, a regime diagram was constructed on the parameter plane of the external forcing (\ref{eq:noisyPulses}), namely the period $T$ and the noise intensity $D$, which serves as an analogue of the forcing amplitude. The corresponding regime diagrams are presented in Fig.~\ref{fig:wgn_uncorr}(a,b) for different ranges of the noise intensity.

As can be seen from the regime diagram, the variety of dynamical regimes is significantly reduced. For small noise intensities $D$, the system still exhibits oscillations around the equilibrium state; however, these oscillations are irregular. For certain initial conditions, a single traveling wave may still be observed. This regime is indicated by the dotted region in the regime diagrams shown in Fig.~\ref{fig:wgn_uncorr}(a,b).

As in the previous section, both regimes coexist with states characterized by multiple traveling waves (Fig.~\ref{fig:wgn_uncorr}(e)). The resulting spatiotemporal patterns appear less smooth compared to the case of purely periodic forcing, which is due to the presence of noise.

At certain noise intensities, a regime resembling spatial synchronization is observed. In this regime, all oscillators exhibit fairly regular dynamics, and their oscillation frequencies are nearly identical. The small mismatch that arises is primarily due to the stochastic nature of the external forcing. The corresponding spatiotemporal diagram is shown in Fig.~\ref{fig:wgn_uncorr}(c). This regime is robust and persists over a broad parameter range, indicated by the orange region in the regime diagrams in Fig.~\ref{fig:wgn_uncorr}(a,b). As the noise intensity increases, a transition from nearly coherent spiking to incoherent dynamics occurs. The regions of existence of coherent spiking resemble synchronization tongues observed for periodic pulse forcing in Fig.~\ref{fig:pulse1}(a).

Analogously to the coexistence of partial spatial synchronization and traveling waves observed under periodic pulsed forcing, Fig.~\ref{fig:wgn_uncorr}(a,b) also reveals the coexistence of nearly coherent oscillations and traveling-wave states (Fig.~\ref{fig:pulse1}(d)). This regime is indicated by the red hatched region.

Due to the stochastic nature of the forcing in (\ref{eq:noisyPulses}), the regime diagrams in Fig.~\ref{fig:pulse1}(a,b) show that a large part of the parameter space is occupied by the green region, where randomly drifting waves are observed. An example of such dynamics is presented in Fig.~\ref{fig:pulse1}(f).

It is also important to note that, in the case of extreme forcing with additive Gaussian white noise, the ensemble does not exhibit such strong multistability as in the case of periodic pulse exposure.

\section{Lévy noise}\label{sec:levy}

In the current paper, we study the coupled FitzHugh-Nagumo oscillators in the excitable regime subject to L{\'e}vy noise. Statistically independent additive L{\'e}vy noise sources $I_i(t)$ in Eqs. (\ref{eq:FHN}) are defined as the formal derivatives of the L{\'e}vy stable motion. L{\'e}vy noise is characterized by four parameters: a stability index $\alpha \in (0:2]$, a skewness (asymmetry) parameter $\beta\in [-1:1]$, a parameter $\mu$ (is assumed to be zero in the current research) being a mean value of the L{\'e}vy noise when $1\leq \alpha \leq 2$ and a scale parameter $s$. The scale parameter can be related with noise intensity $D=\sigma^{\alpha}$.
%The characteristic function of noise sources $\xi_i(t)$ takes the form \cite{janicki1994,dybiec2006,dybiec2007}:
To generate L{\'e}vy noise signals $\eta_t$, the Janicki--Weron algorithm was applied for the fixed parameter values $\beta=\mu=0$ and a variable scale factor $1<\alpha \leq 2$. In such a case, the formula for random number generation takes the simplified form (see papers \cite{janicki1994,weron1995} for more details):
\begin{equation}
\label{eq:noise_generation} 
I(t)=\eta_t = s \times \dfrac{\sin(\alpha V_t)}{(\cos(V_t))^{1/\alpha}}\times \left( \dfrac{\cos(V_t(1-\alpha))}{W_t}\right)^{\dfrac{1-\alpha}{\alpha}}, 
\end{equation} 
where $V_t$ are random variables uniformly distributed on $\left(-\dfrac{\pi}{2}:\dfrac{\pi}{2}\right)$, $W_t$ are exponential random sequences with mean 1 (variables $W$ and $V$ are statistically independent). A similar numerical procedure was used in Ref. \cite{korneev2024} addressing the issue of L{\'e}vy noise-induced coherence resonance in the single FitzHugh--Nagumo oscillator in the excitable regime. In case $\alpha=2$, signals $I(t)$ represent independent sources of white Gaussian noise. In case $\alpha<2$, the noise distribution is non-Gaussian and contains long heavy tails associated with random impulses of high amplitude.

To analyze the effect of Lévy noise on the FitzHugh--Nagumo ensemble, the integration step was reduced to $h = 10^{-4}$. For simulations with large values of $s$, the step size had to be decreased further, since large-amplitude random jumps induced by Lévy noise can drive the trajectory to infinity if the integration step is chosen too large.

\begin{figure}[t]
\includegraphics[width=\linewidth]{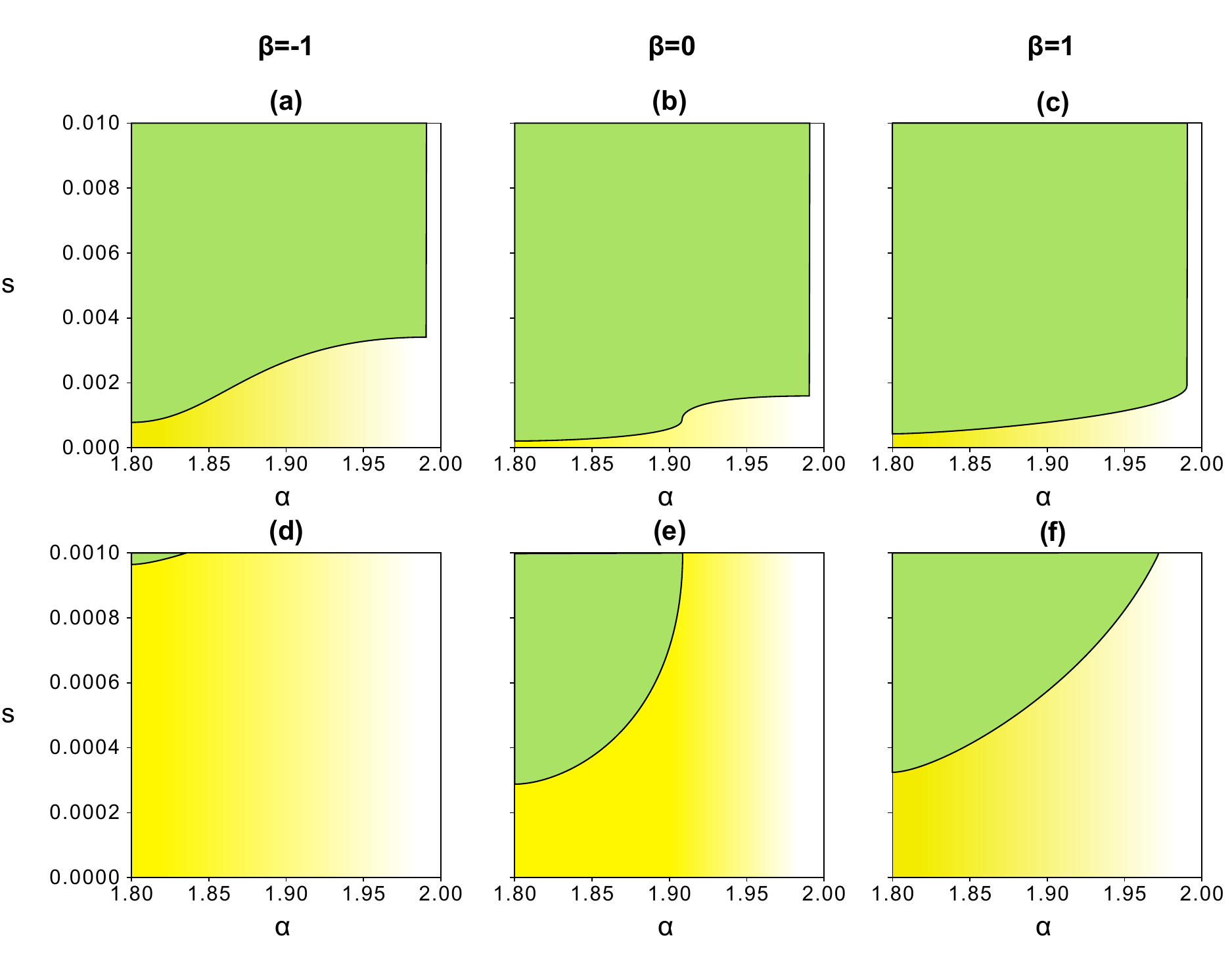}
\caption{\label{fig:levy} Regime diagram on the parameter plane $(s, \alpha)$ of Lévy noise with $\beta=-1$ (a,d), $\beta=0$ (b,e) and $\beta=1$ (c,f). Yellow region correspond to weak oscillations around the equilibrium state with random single spikes. This regime coexists with single traveling wave and two simultaneously excited waves propagating in opposite directions. The last regime can be also observed in the green area without coexistence of other regimes.}
\end{figure}

Figure~\ref{fig:levy} shows regime diagrams on the parameter plane of the noise parameters $s$ and $\alpha$ for three different values of $\beta = 0$, $\beta = \pm 1$, corresponding to a symmetric case and two asymmetric probability distributions with shifted means, respectively. To construct the regime diagrams, the ranges of the parameter $s$ were taken up to $10^{-2}$ (upper panels (a--c)) and up to $10^{-3}$ (lower panels (d--f)).

For $\alpha = 2$, the Lévy noise becomes white Gaussian noise; therefore, in all regime diagrams the case $\alpha = 2$ corresponds either to a steady state with weak oscillations around the equilibrium state or to a single traveling wave (Fig.~\ref{fig:trav_wave}), depending on the initial conditions. As the parameter $\alpha$ decreases, the specific properties of Lévy noise lead to the emergence of increasingly large-amplitude excursions. As a result, in all regime diagrams shown in Fig.~\ref{fig:levy}, the yellow area represents the regime in which most oscillators exhibit small unregular oscillations around the equilibrium state, while sporadic spike-like bursts occur randomly throughout the ensemble. Moreover, this regime is represented using a gradient, since as $\alpha$ approaches 2, the amplitude of these bursts decreases continuously and the excursions vanish completely at $\alpha = 2$.

On all regime diagrams, this state coexists with the regime of a single traveling wave, as well as with the regime in which two simultaneously excited waves propagate in opposite directions, subsequently collide, and annihilate. As the scale parameter increases, such waves are excited in the ensemble more frequently, and with increasing $s$ the dynamics become increasingly similar to random wandering. The region where neither small oscillations around the equilibrium nor a single traveling wave are observed, and where only either two simultaneously excited counter-propagating waves or, at larger $s$, random walking dominates, is shown in green in the regime diagrams in Fig.~\ref{fig:levy}.

\section{Conclusion}\label{sec:conclu}

The present work investigates the influence of extreme external forcing on the emergence of traveling waves in an ensemble of excitable FitzHugh--Nagumo systems couples with weak nonlocal coupling. As representative examples of such forcing, we consider a periodic sequence of Gaussian pulses, the same pulse sequence multiplied by a stochastic modulation with Gaussian white noise, and Lévy noise with different parameter settings controlling the symmetry of the probability distribution and its mean value. It is shown that periodic forcing in both cases leads to the appearance of synchronization tongues, within which all oscillators in the ensemble exhibit highly regular dynamics with nearly identical frequencies. Under periodic excitation, multiple co-propagating traveling waves may also form. In addition, in the presence of temporal periodicity in the external input, regimes can arise in which partial spatial synchronization coexists with traveling-wave dynamics.

Lévy noise does not lead to regular dynamics; however, it gives rise to a new regime in which two waves propagating in opposite directions are simultaneously excited. This regime exists over a wide parameter range, but as the scale parameter of the noise increases, such wave excitations occur more frequently, and the dynamics eventually transform into random walking behavior.

The present study is focused on a detailed characterization of all emerging regimes, together with their representation on parameter maps of the extreme external forcing.

\begin{acknowledgments}
This work was supported by the Russian Science Foundation (project No. 23-72-10040). %\hyperref{https://rscf.ru/project/23-72-10040/}
\end{acknowledgments}

\section*{Data Availability Statement}
The data that support the findings of this study are available from the corresponding author upon reasonable request.

\bibliography{bibliography}

\end{document}